\documentclass[aps,prd,secnumarabic,nobibnotes,twocolumn,superscriptaddress]{revtex4-1}
\usepackage{amsfonts}
\usepackage{mathrsfs}
\usepackage{amsmath}
\usepackage{color}
\usepackage{natbib}
\usepackage{graphicx}
\usepackage{bm}
\usepackage{amssymb}
\usepackage{xspace}
\usepackage{epstopdf}
\usepackage{dcolumn}
\usepackage{multirow}
\usepackage[colorlinks=true, letterpaper=true, pdfstartview=FitV, linkcolor=blue, citecolor=blue, urlcolor=blue]{hyperref}
\usepackage{wrapfig}

\makeatletter

\newcommand{\Rmnum}[1]{\expandafter\@slowromancap\romannumeral #1@}
\makeatother

\begin{document}
\title{Ideal inner nodal chain semimetals in Li$_2$XY (X = Ca, Ba; Y = Si, Ge) materials }

\author{Xiaoming Zhang}
\address{State Key Laboratory of Reliability and Intelligence of Electrical Equipment, Hebei University of Technology, Tianjin 300130, China.}
\address{School of Materials Science and Engineering, Hebei University of Technology, Tianjin 300130, China.}

\author{Lei Jin}
\address{School of Materials Science and Engineering, Hebei University of Technology, Tianjin 300130, China.}

\author{Xuefang Dai}
\address{School of Materials Science and Engineering, Hebei University of Technology, Tianjin 300130, China.}

\author{Guifeng Chen}
\address{School of Materials Science and Engineering, Hebei University of Technology, Tianjin 300130, China.}

\author{Guodong Liu}\email{gdliu1978@126.com }
\address{State Key Laboratory of Reliability and Intelligence of Electrical Equipment, Hebei University of Technology, Tianjin 300130, China.}
\address{School of Materials Science and Engineering, Hebei University of Technology, Tianjin 300130, China.}

\begin{abstract}
The chain-type nodal loops in the reciprocal space can generate exotic nodal chain fermions. Here, we report that Li$_2$XY (X = Ca, Ba; Y = Si, Ge) compounds are ideal inner nodal chain semimetals. Their band structures are composed of two connecting nodal loops with either hybrid or type-I band dispersion. The signatures of the nodal chain, such as the nontrivial surface states, are quite pronounced in these Li$_2$XY compounds since there is only a single inner nodal chain without other extraneous bands near the Fermi level. These compounds are existing materials and ambient-stable, which is available to realize the experimental detection of inner nodal chain fermions or further the practical applications.
\end{abstract}
\maketitle

Motivated by the discoveries of Weyl~\cite{add1,add2,add3,add4,add5} and Dirac semimetals~\cite{add6,add7,add8,add9,add10}, topological semimetals currently attract significant research interests in condensed matter physics. Because of the existence of nontrivial band-crossings near the Fermi energy, topological semimetals can show exotic transport, magnetic and optical phenomena and great potential in electronics and quantum computing applications~\cite{add11,add12,add13,add14,add15}. The band-crossings in topological semimetals can not only form zero-dimensional (0D) nodal points (including Weyl, Dirac and three-, six-, eight-fold degenerated nodal points~\cite{add16}), but also can lead to one-dimensional (1D) nodal loop~\cite{add17,add18,add19} and two-dimensional (2D) nodal surface~\cite{add20,add21,add22,add23}. Nodal loop semimetal has been proposed in many realistic materials so far~\cite{add17,add18,add19,add24,add25,add26,add27,add28,add29,add30,add31,add32,add33,add34,add35,add36,add37,add38,add39}. Many interesting properties including the drumhead surface state~\cite{add17,add19}, the anisotropic transport~\cite{add40,add41,add42,add43}, the unconventional optical response~\cite{add44,add45}, and the potential surface magnetism and high-temperature superconductivity~\cite{add46,add47,add48,add49} were reported.
	
Compared with nodal point semimetal, nodal loop semimetal possess more variety of nodal structures. Such variety could be either from the dispersion around the band crossing or from its configuration. For the former, nodal loop is not limited to conventional type-I and type-II, but possesses the third occasion of the coexistence of type-I and type-II points in one loop (hybrid nodal loop). Hybrid nodal loop semimetal was notified recently~\cite{add50,add51,add52}, and only quite limited candidate materials were proposed~\cite{add52}. Regarding the latter, nodal loop can take various configurations including a single loop, multiple crossing loops~\cite{add53,add54}, nodal chains~\cite{add55,add56,add57}, Hopf links~\cite{add58,add59,add60,add61,add62}, and etc. Nodal chain semimetal was initially proposed by Bzdu\"{S}ek \emph{et al.} in IrF$_4$ materials~\cite{add55}, and by Yu \emph{et al.} in HfC materials~\cite{add63}, along with anomalous magnetotransport properties. Later on, Chang \emph{et al.} further classified nodal chain into outer and inner nodal chain, and predicted their coexistence in Heusler Co$_2$MnGa material~\cite{add60}. Recently, the outer nodal chain state has been observed by Yan \emph{et al.} in a metallic-mesh photonic crystal~\cite{add64}. However, there is no experimental verification of an inner nodal chain semimetal yet, mostly because the electronic band structures of previous materials suffer from various drawbacks, such as the nodal chain situates away from the Fermi level, their exist other extraneous bands near the nodal chain, and so on. Therefore, ideal candidate materials are needed. As an ideal inner nodal chain semimetal, the candidate material needs to at least satisfy the following requirements. First, the nodal chain should be near the Fermi level. Second, the nodal chain needs to have a relatively simple shape, and it is the best that the system possesses a single chain composed of two, only two, closed loops. Third, it is crucial that near the nodal chain there is no extraneous bands, since the extraneous bands will strongly interfers with the unique physical properties related to the nodal chain semimetal. Fourth, the candidate materials should be stable and easy to be synthesized and facilitates the experimental studies. These rigorous conditions limit the development of suitable inner nodal chain candidates, and it is urgent to search for realistic materials that satisfy these requirements.
	
In current work, based on first-principles calculations and symmetry analysis, we propose that, a family of ambient-stable materials, lithium alkaline earth tetrelides Li$_2$XY (X = Ca, Ba; Y = Si, Ge), are ideal inner nodal chain semimetals that satisfy all the above mentioned requirements. Taking Li$_2$BaSi as a concrete example, we show that the material features a single inner nodal chain near the Fermi energy in the band structure. The nodal chain consists of two perpendicular closed nodal loops that situate in non-equivalent mirror/glide-mirror planes. Beside the two bands that form the nodal chain, there is no other extraneous bands nearby. As a result, clear drumhead surface states from the nodal chain are observed. More interestingly, we find that the two nodal loops of nodal chain in Li$_2$BaSi are both hybrid loops where type-I and type-II nodal points coexist. Therefore, our prediction provides ideal candidate materials for experimentally exploring the novel properties of inner nodal chain and hybrid nodal loop states.

\begin{figure}
\includegraphics[width=8.8cm]{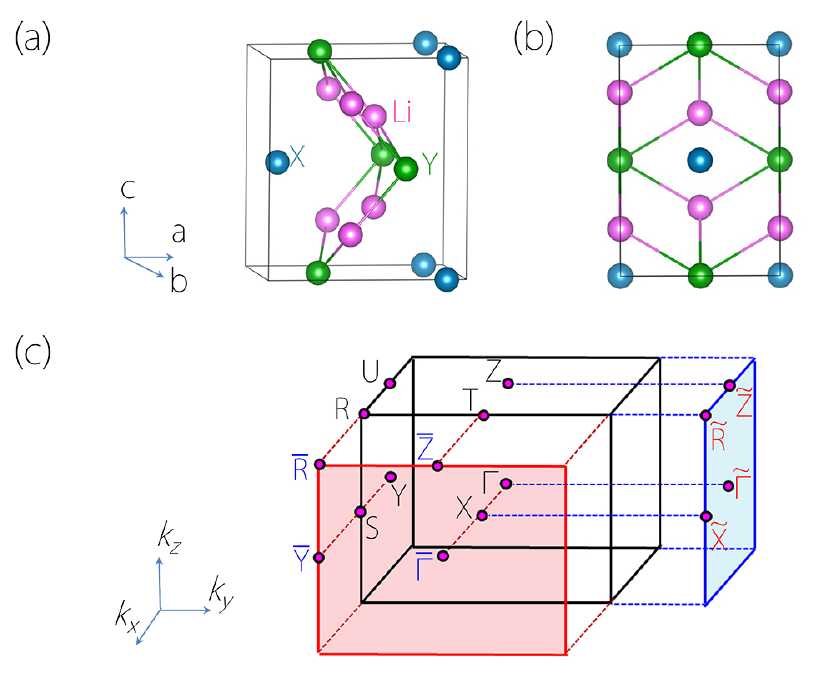}
\caption{(a) Crystal structure of ternary Li$_2$XY (X = Ca, Ba; Y = Si, Ge) compounds. (b) is the side view of the crystal structure. (c) The bulk Brillouin zone and its projection onto the (010) and (100) surfaces. The high-symmetry points are labelled.
\label{fig1}}
\end{figure}

The family of Li$_2$XY (X = Ca, Ba; Y = Si, Ge) compounds are all existing materials and stable under ambient condition. Especially, their high-quality single crystal samples have been synthesized by the flux method~\cite{add65,add66}. They naturally crystallize in an orthorhombic structure, with the space group \emph{Pmmns} (No. 59). As shown in Fig.~\ref{fig1}(a) and \ref{fig1}(b), the bonding between Li and Y atoms forms buckled sheets along the \emph{z} direction (\emph{c}-axis), and X atoms make up the framework of the orthorhombic lattice. In the unit cell, Li, X and Y atoms occupy the 4\emph{e} (0, \emph{u}, \emph{v}), 2\emph{a} (0, 0, \emph{z$_a$}) and 2\emph{b} (0, 0.5, \emph{z$_b$}) Wyckoff sites, respectively. The crystal structure possesses the following important symmetries: the inversion \emph{P}, the mirror reflections\emph{ m$_x$} and \emph{m$_z$}, and the glide-mirror symmetry \emph{ g$_y$}: (\emph{x}, \emph{y}, \emph{z}) ¡ú (\emph{x}+1/2, \emph{y}+1/2, -\emph{z}). In addition, the materials show no magnetic ordering, so the time reversal symmetry \emph{T} is also preserved in the system.

\begin{figure}
\includegraphics[width=8.8cm]{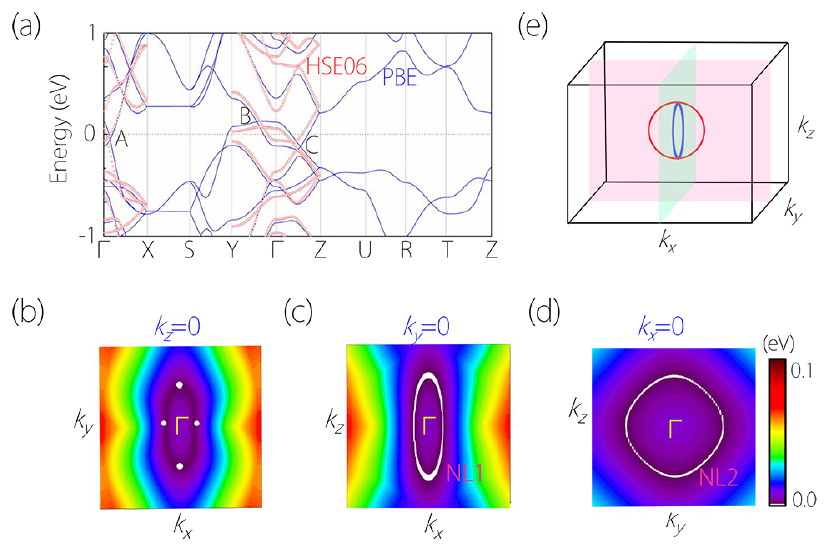}
\caption{(a) Calculated electronic band structure of Li$_2$BaSi compound with PBE (blue lines) and HSE06 (red dots). The shape of possible nodal loops in (b) \emph{k$_z$}=0 plane, (c) \emph{k$_y$}=0 plane, and (d) \emph{k$_x$}=0 plane. The colour-map shows the local gap between the conduction and valence bands. (e) The schematic illustration of the inner nodal chain in Li$_2$BaSi.
\label{fig2}}
\end{figure}

First-principles calculations are performed to study the material properties, and the computational details are shown in the Supplementary Material~\cite{add67}. The electronic band structure of Li$_2$BaSi without spin-orbit coupling (SOC) is shown in Fig.~\ref{fig2}(a). Both calculated results by Perdew-Burke-Ernzerhof (PBE, see the blue lines) and Heyd-Scuseria-Ernzerhof (HSE06, see the red dots) hybrid functionals give out semi-metallic band structures, which is consistent with the experimental results~\cite{add65}. In the band structures, we can observe three band-crossing points near the Fermi level: crossing-A on $\Gamma$-A path, crossing-B on $\Gamma$-Y path, and crossing-C on $\Gamma$-Z path. Because of the presence of both \emph{P} and \emph{T} symmetries, the band-crossing points in Li$_2$BaSi cannot be isolated~\cite{add17}. Noticing the three band-crossing points belong to \emph{k$_z$}=0, \emph{k$_y$}=0 or \emph{k$_x$}=0 planes, we make a careful scan of band structures on these planes. As shown in Fig.~\ref{fig2}(b), \ref{fig2}(c) and \ref{fig2}(d), we indeed find these points belong to two closed nodal loops centering the $\Gamma$ point: one resides in \emph{k$_y$}=0 plane [denoted as NL1 in Fig. 2(c)] and the other situates in \emph{k$_x$}=0 plane [denoted as NL2 in Fig.~\ref{fig2}(d)]. One also note that, there is no nodal loop in \emph{k$_z$}=0 plane [see Fig.~\ref{fig2}(b)].

To further characterize the nodal loops, we construct a low-energy effective Hamiltonian around the $\Gamma$ point. Given by the symmetry analysis, the conduction and valence bands in Li$_2$BaSi correspond to irreducible representations \emph{B$_{2u}$} and \emph{B$_{3u}$} of the \emph{D$_{2h}$} symmetry, respectively. Taking them as basis states, the 2$\times$2 effective Hamiltonian for the two bands around the $\Gamma$ point takes the general form:
\begin{equation}\label{FNRm}
\mathcal{H}=\left[
              \begin{array}{cc}
                M_{1}+A_{1}k_{x}^{2}+B_{1}k_{y}^{2}+C_{1}k_{z}^{2} & Dk_{x}k_{y} \\
                Dk_{x}k_{y} & M_{2}+A_{2}k_{x}^{2}+B_{2}k_{y}^{2}+C_{2}k_{z}^{2} \\
              \end{array}
            \right],
\end{equation}
Here, the expansion is up to \emph{k}-quadratic terms, and \emph{M$_i$}, \emph{A$_i$}, \emph{B$_i$} and \emph{C$_i$} with i=1,2, and \emph{D} are material-specific coefficients. The Hamiltonian indicates that, a band-crossing point in \emph{k$_x$}=0 (and \emph{k$_y$}=0) plane will produce a nodal loop, which describes well the DFT results.

In fact, both nodal loops are protected by two independent symmetries when SOC is absent. One is the coexistence of \emph{P} and \emph{T} symmetries, which requests the Berry phase for any close path to be quantized in unit of $\pi$. We numerically calculated the Berry phase for each loop, the result give out to be $\pm$$\pi$, ensuring the protection of them from opening gap against weak perturbations. The other protection is the glide-mirror symmetry \emph{g$_y$} (mirror symmetry \emph{m$_x$}) for NL1 (NL2). This requires the conduction and valence bands in Li$_2$BaSi possess opposite \emph{g$_y$} (\emph{m$_x$}) eigenvalues in the glide-mirror (mirror) plane \emph{k$_y$}=0 (\emph{k$_x$}=0). which have been verified by our DFT calculations. The \emph{g$_y$} (\emph{m$_x$}) symmetry protects the NL1 (NL2) to exactly occur within \emph{k$_y$}=0 (\emph{k$_x$}=0) plane.

Interestingly, the two nodal loops are not isolated, but touch with each other at crossing-C, as shown in Fig.~\ref{fig2}(e). As we have mentioned, nodal chain structure can be classified into two types (outer nodal chain and inner nodal chain) based on the connecting configuration of the nodal loops. In an outer nodal chain, the two nodal loops are on the opposite sides of touching point, while the loops situate on the same side in the inner nodal chain. It is also worth noticing that, although possesses similar looking, the inner nodal chain is fundamentally different from traditional crossing nodal loops, as shown in Fig. S1 of the Supplemental Material~\cite{add67}: traditional crossing nodal loops are equivalent with each other and are protected by the the same mirror symmetry~\cite{add54}, while the loops in inner nodal chain are non-equivalent. In the current case, from Fig.~\ref{fig2}(e), one can easily find that the nodal loops in Li$_2$BaSi forms the inner nodal chain structure.

\begin{figure}
\includegraphics[width=8.8cm]{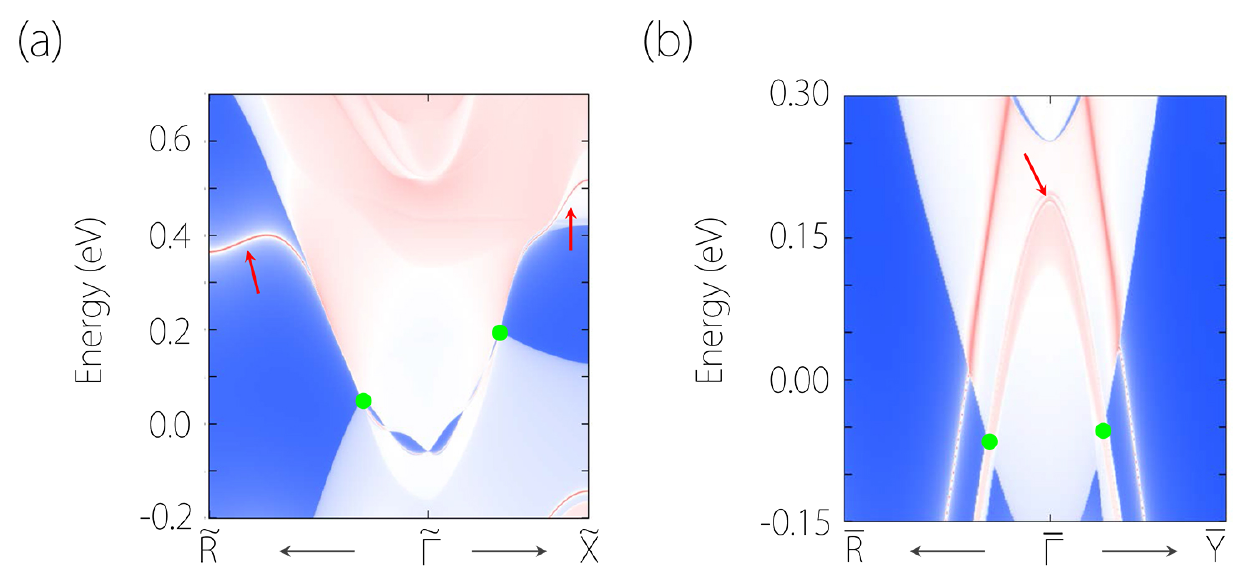}
\caption{(a) Projected spectrum on the (010) surface of Li$_2$BaSi compound. (b) is similar with (a) but for the (100) surface projection. The termination of the (010) and (100) surfaces is Li-Ba and Si, respectively. In (a) and (b), the green dots mark the projected bulk band-crossing points on the nodal loop, and the red arrows point the drumhead surface states.
\label{fig3}}
\end{figure}

Nodal loops usually manifest drumhead surface states~\cite{add17}. In Li$_2$BaSi, the two nodal loops of the nodal chain situate in \emph{k$_y$}=0 and \emph{k$_x$}=0 planes, respectively. Thus we calculated the spectra for both (010) and (100) surfaces, as shown in Fig.~\ref{fig3}(a) and \ref{fig3}(b). The nontrivial surface bands emanated from the bulk nodal points are indeed observed for both surface projections. In Li$_2$BaSi compound, the existence of drumhead surface states in non-equivalent surfaces can facilitate the experimental detection of the surface bands by surface-sensitive techniques.

\begin{figure}
\includegraphics[width=8.8cm]{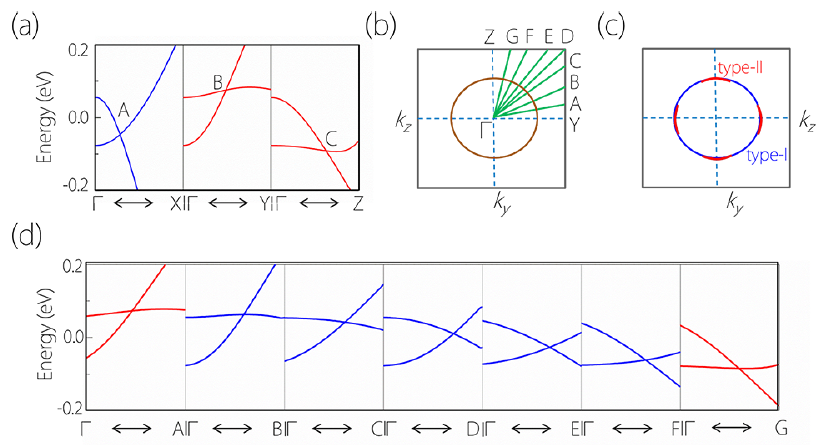}
\caption{(a) Enlarged view of HSE06 band structures around crossing-A, -B, and -C for Li$_2$BaSi compound. (b) The shape of NL2 in \emph{k$_x$}=0 plane, and selected \emph{k} paths through NL2. (d) Band structures along the \emph{k} paths indicated in (b). In (a) and (d), the bands with type-I and type-II dispersions are shown in blue and red colours, respectively.
\label{fig4}}
\end{figure}

After a careful investigation on the band structure of Li$_2$BaSi achieved by HSE06 calculation [see Red dots in Fig.~\ref{fig2}(a)], we find the band dispersion around crossing-A, -B, and -C has different slopes. As shown in the enlarged band structure in Fig.~\ref{fig4}(a), one clearly finds that crossing-A is a type-I nodal point, while crossing-B and -C are type-II ones. Noticing crossing-A and -C belong to NL1, NL1 is in fact a hybrid type nodal loop with the coexistence of type-I and type-II nodal points. For NL2, beside crossing-B and -C points (both type-II), we carefully examine the band dispersion for other points on the loop. As shown in Fig.~\ref{fig4}(b) and \ref{fig4}(c), there coexist type-I and type-II points on NL2, which shows the hybrid nodal loop signature of NL2. Therefore, the nodal chain in Li$_2$BaSi consists of two hybrid nodal loops. Considering hybrid nodal loop state has been rarely reported in realistic materials (the only report is in Ca$_2$As~\cite{add52}), Li$_2$BaSi compound can be another excellent candidate as a hybrid nodal loop material.

Before closing, we have two remarks. First, As shown in the Supplementary Material~\cite{add67}, the inner nodal chain signature is also found in related materials with the same lattice structure, including Li$_2$CaSi and Li$_2$CaGe, although the nodal chains in these materials are only composed of type-I nodal loops. Here, our report provides more choice to study the properties of nodal chain state, as well as the differences between hybrid and type-I nodal loops. Second, SOC is neglected in our above discussions. Although SOC generally gaps the nodal loops, the SOC-induced gaps are very small because of the small SOC strength in these materials. Our computations reveal that the gap-size is only 18-1.7 meV in these materials (see the Supplementary Material~\cite{add67}), which is small enough to be neglected. So the inner nodal chain state and its exotic properties are achievable in the Li$_2$XY (X = Ca, Ba; Y = Si, Ge) materials.

In conclusion, by using Li$_2$BaSi as a concrete example, we propose that the Li$_2$BaSi family materials are ideal inner nodal chain semimetals. There is a single nodal chain composed of two perpendicular nodal loops in the band structure close to the Fermi energy. Near the nodal chain, there is no other extraneous band. The drumhead surface states corresponding to the nodal chain are clearly observed. Such an ideal inner nodal chain semimetal has not been identified before. The nodal loops in the nodal chain are very robust due to under the protection of two independent mechanism (either \emph{PT} symmetry or mirror/glide-mirror symmetry). An effective Hamiltonian is constructed to describe the nodal chain. Moreover, we find the nodal chain in these materials makes up either from hybrid or type-I nodal loops. This family of materials are stable and quite easy to synthesize, which further make them as suitable candidate materials to investigate the interesting physics associated with nodal chain, as well as type-I and hybrid nodal loops.¡°\\¡±

This work is supported by the Special Foundation for Theoretical Physics Research Program of China (No. 11747152), Chongqing City Funds for Distinguished Young Scientists (No. cstc2014jcyjjq50003). One of the authors (G.D. Liu) acknowledges the financial support from Hebei Province Program for Top Young Talents.

\end{document}